\documentclass[pdflatex,sn-mathphys-num]{sn-jnl}

\usepackage{graphicx}
\usepackage{multirow}
\usepackage{amsmath,amssymb,amsfonts}
\usepackage{amsthm}
\usepackage{mathrsfs}
\usepackage[title]{appendix}
\usepackage{xcolor}
\usepackage{textcomp}
\usepackage{manyfoot}
\usepackage{booktabs}
\usepackage{algorithm}
\usepackage{algorithmicx}
\usepackage{algpseudocode}
\usepackage{verbatim}
\usepackage{listings}

\usepackage{multicol}
\usepackage{tikz}

\begin{document}

\title[Article Title]{Machine-Learned Force Fields for Lattice Dynamics at Coupled-Cluster Level Accuracy}

\author*[1]{\fnm{Sita} \sur{Sch\"onbauer}}\email{sita.schoenbauer@tuwien.ac.at}

\author[2,1]{\fnm{Johanna P.} \sur{Carbone}}

\author[1]{\fnm{Fredrik V.} \sur{Eriksson}}

\author[1]{\fnm{Florian} \sur{Libisch}}

\author[1]{\fnm{Andreas} \sur{Gr\"uneis}}

\affil*[1]{\orgdiv{Institute of Theoretical Physics}, \orgname{Technical University of Vienna}, \orgaddress{\street{Wiedner Hauptstra\ss e 8–10}, \postcode{1040} \city{Vienna}, \country{Austria}}}

\affil[2]{\orgdiv{Faculty of Physics and Center for Computational Materials Science}, \orgname{University of Vienna}, \orgaddress{\street{ Kolingasse 14-16}, \postcode{1090} \city{Vienna},  \country{Austria}}}

\abstract{
We investigate Machine-Learned Force Fields (MLFFs) trained on
approximate Density Functional Theory (DFT) and Coupled Cluster (CC) level 
potential energy surfaces for the carbon diamond and lithium hydride solids.
We assess the accuracy and precision of the MLFFs by calculating phonon dispersions and vibrational densities of states (VDOS) that are compared to experiment and reference ab initio results.
To overcome limitations from long-range effects and the lack of atomic forces in the CC training data, a delta-learning approach based on the difference between CC and DFT results, as well as a charge aware MLFF approach is explored.
Compared to DFT, MLFFs trained on CC theory yield higher vibrational frequencies for optical modes, agreeing better with experiment.
Furthermore, the MLFFs are used to estimate anharmonic effects on the VDOS of lithium hydride at the level of CC theory.

}

\keywords{Machine Learning, Phonons, Coupled Cluster, Density Functional Theory}

\maketitle

\begin{multicols}{2}

\newpage

\section{Introduction}\label{intro}

Machine learning (ML) methods have become an invaluable asset in determining precise atomic force fields without having to explicitly keep track of the electronic problem~\cite{mlmd1,mlmd2gpr,mlmd3,mlmd4code,mlmd5symm,Unke2021,behler}. This allows for increases in system size and, in the case of Molecular Dynamics (MDs), longer simulation times at comparatively negligible computational cost.
However, the predictive accuracy of ML models is inherently limited by the quality of the training data used to generate them.

So far, most ML studies in the field of materials science are limited to data
generated using approximate exchange and correlation density
functionals.
More accurate and systematically improvable Wave Function Theory (WFT) based methods such as
the ``gold standard" in quantum chemistry, Coupled Cluster (CC) theory at the level of single, double and perturbative triple particle-hole excitations (CCSD(T)) are computationally significantly more expensive, which severely limits the size of accessible training data~\cite{Bartlett2007,Raghavachari1989}.
Consequently, these methods have been used mainly to generate training data for molecules~\cite{Ruth2023}.
Recently, several studies have expanded their scope in the
context of ML to calculate molecular adsorption energies in a zeolite structure and the properties of water~\cite{Daru2022,chen2023data,herzog2024coupled}.
In this work, we investigate the extension of ML approaches to study the lattice dynamics of solids from training
data generated by periodic CCSD(T) theory.

Phonons play a fundamental role in determining various materials properties, including heat capacity and electron-phonon coupling, which is closely related to the emergence of exotic states of matter such as superconductivity~\cite{superconductivity,kulic2000interplay} and quantum paraelectricity~\cite{paraelectricity,ranalli2023temperature}. Therefore, efficient and accurate computational methods for simulating phonons are essential for designing quantum materials with tailored and enhanced properties. In this context, DFT persuades with its great cost-accuracy trade-off for many materials.
However, currently available density functionals sometimes
fail to achieve the required level of accuracy.

A specific example are optical phonon frequencies in diamond, which are underestimated by commonly used approximate density functionals based on the generalized gradient approximation~\cite{cqmcgamma}. 
Yet, highly accurate knowledge of the relationship between phonon
frequency and pressure is needed for diamond anvil cell experiments~\cite{cqmcgamma,cgammaexp}.

Training data containing energies and atomic forces substantially improves the trade-off between precision of the MLFF and the required training data size.
However, there exist ab initio methods for which the calculation of atomic forces is difficult, 
because it requires the implementation of many terms, which might introduce excessive usage of memory.
Therefore, it is also desirable to investigate approaches that are based on training data that exclude atomic forces.
Currently, $\Delta$- and transfer-learning approaches provide such an avenue~\cite{deltacc1,deltacc2,Ruth2023,transfer,transfer2,transfermariia}, and there have also been promising results getting to CCSD(T) accuracy without forces~\cite{Daru2022}.
Here, we restrict our investigation to a single $\Delta$-learning 
technique for periodic CCSD(T) theory because a larger number
of different ML approaches would be beyond
the scope of this work. Moreover, the number of systems studied
is mostly limited by the relatively large computational cost of CCSD(T) calculations.

Many successful approaches for MLFFs are described in literature.
In particular, neural networks and regression models have become very popular in materials science during the last decades~\cite{Behler2007,Bartok2010}.
We employ an equivariant message passing neural network referred to as \texttt{MACE}~\cite{Batatia2022mace,Batatia2022Design}.
The \texttt{MACE} implementation requires few input parameters,
enabling a straightforward application to training data generated by periodic CC theory. Additionally, \texttt{QNEP} \cite{qnep1,qnep2,qnep3} is used to bridge the gap to obtain long-range forces in the ionic LiH crystal; however, it is only trained at a DFT level used as the base for the $\Delta$-learning approach.

The main goals of the present work are to: (i) further assess the potential
of $\Delta$-learning approaches for periodic CCSD(T) calculations, (ii) contribute to directing
future developments in periodic CC theory needed to take full advantage of MLFFs,
and (iii) establish a workflow that can be used in future applications of periodic CC calculations, where MLFFs are needed to bridge time- and system-size scales currently limited by the computational cost.
The training data generated for the present study is made publicly available to serve as input data for other techniques.

\section{Results}\label{res}

We employ the following notation throughout. ML($X$) refers to predictions made by an MLFF (\texttt{QNEP} or \texttt{MACE}) trained on data generated using method $X$. In the case of $X=\text{DFT}_\text{E}$, the training data includes energies computed using an approximate exchange and correlation density functional only.  $X=\text{DFT}_\text{E,F}$ also includes atomic forces in the training data. $\Delta$ML($X$) refers to results generated using a simple linear combination of two MLFFs. The first MLFF corresponds to ML(DFT$_\text{E,F}$), whereas the second MLFF is trained on the difference in energy between WFT method $X$ and DFT. This includes Hartree-Fock (HF), M\o ller-Plesset perturbation theory (MP2), CC singles and doubles (CCSD) and CC singles, doubles, and perturbative triples (CCSD(T)). More details are explained in Sec.~\ref{meth}.

\subsection{\label{c}Diamond}

\begin{figure*}[t]
  \includegraphics[width=\textwidth]{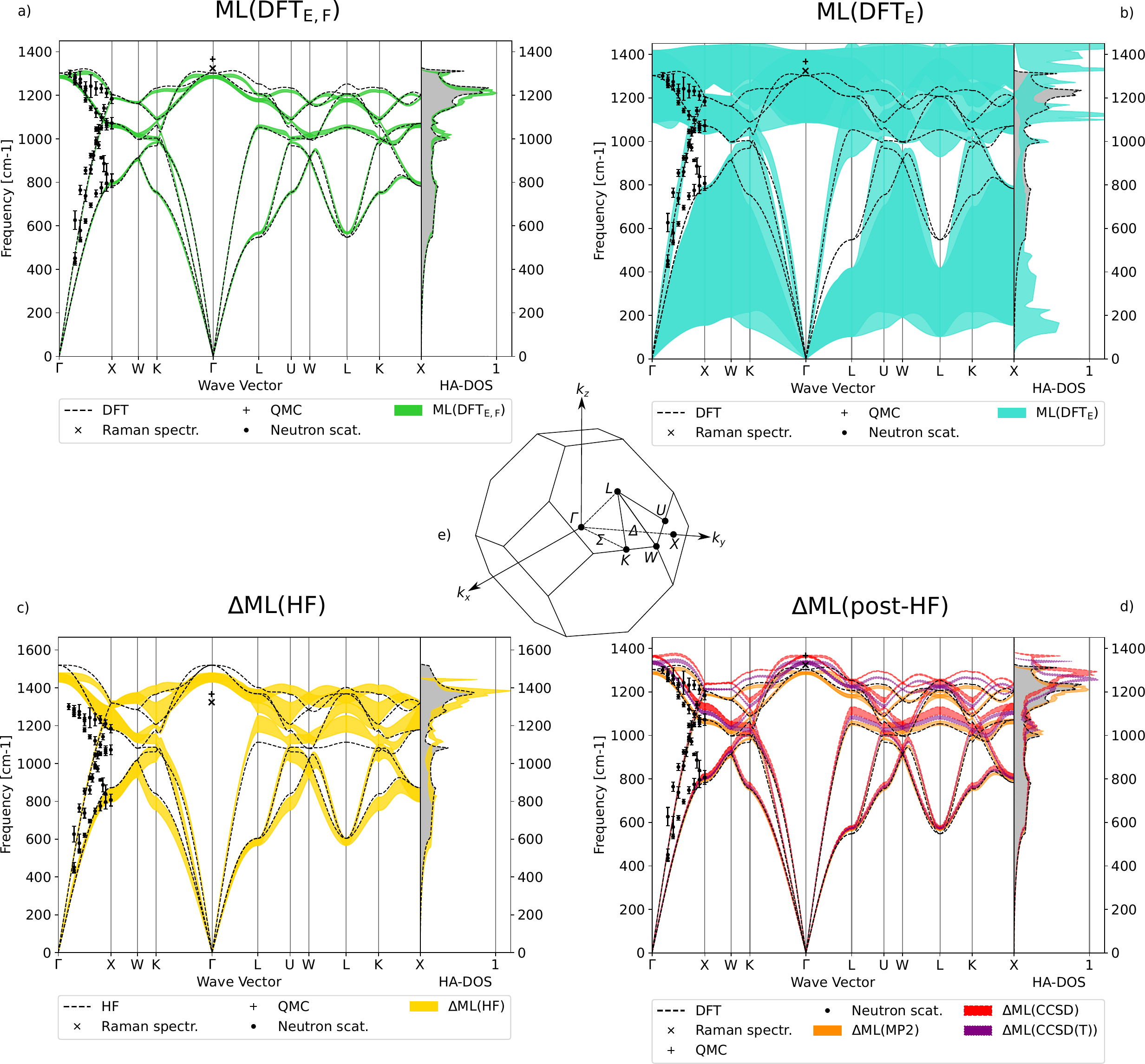}
  \caption{Phonon dispersions and DOS comparisons for diamond; neutron scattering data (black dots) from~\cite{cexp}, Raman spectroscopy (black x) from~\cite{cgammaexp} and quantum Monte Carlo data (black plus) from~\cite{cqmcgamma}. a) ML(DFT$_{\rm{E,F}}$) (green area) vs. DFT results (black dashed line). b) ML(DFT$_{\rm{E}}$) (trained only on E) (cyan area) vs. DFT results (black dashed line). c) $\Delta$ML(HF) (yellow area) vs. HF results (black dashed line). d) Different $\Delta$ML(WFT) results (gold, red dashed, purple dotted areas) vs. DFT results (black dashed line). e) Brillouin zone with relevant high symmetry points for diamond.}
  \label{fig:cphon}
\end{figure*}

We start by discussing the phonon dispersion of carbon diamond.
First, we focus on results generated using a MLFF trained on $200$ training data points of DFT-PBE ground-state energies and forces. These data points have been randomly chosen from an MD trajectory, as described in Sec.~\ref{ccomp}.
Fig.~\ref{fig:cphon} a) shows that ML(DFT$_{\rm{E,F}}$) (green area) is able to satisfactorily reproduce the DFT phonon dispersions and harmonic approximation density of states (HA-DOS) (black dashed line).

The green area depicts the maximum and minimum phonon
frequency obtained for 5 different random seeds used
in the training of the MLFFs. Note that the same randomly
selected data points have been used during training.
We refer to this disagreement as ``seed based spread''.
For a depiction of each individual seed, see Sec. 1.3 of the SI~\cite{si}.

All seeds strongly overlap with the DFT results (black dashed line), except 
for at the L point,
where ML(DFT$_{\rm{E,F}}$) underestimates the upper optical and lower acoustic mode 
frequencies.
This can be understood by noting that the MLFFs are trained on a 
$2\times2\times1$ supercell of the conventional cell, where the L 
point is not represented (see Fig.~\ref{fig:cphon} e) - the L 
point points furthest ``up").
Due to the relatively large computational cost of CCSD(T) 
calculations, larger supercells could not be studied.

Next, we inspect an MLFF trained only on DFT energies and compare it with reference DFT results. In Fig.~\ref{fig:cphon} b), it is clear that ML(DFT$_{\rm{E}}$) (cyan area) cannot even qualitatively reproduce the phonon dispersions of DFT and exhibits notably larger seed-based spread, clearly demonstrating
the need for atomic forces in the training data. 

As a possible way to avoid the need for atomic forces for beyond-DFT methods, we explore a $\Delta$-learning approach (described in detail in Sec.~\ref{sec:delta}). To verify that $\Delta$-learning is able to reproduce phonon dispersions of the respective parent method, we look at $\Delta$ML(HF), shown in Fig.~\ref{fig:cphon} c) (yellow area), as compared to a phonon dispersion at the level of HF theory (black dashed line). In contrast to ML(DFT$_{\rm{E,F}}$) shown in Fig.~\ref{fig:cphon} a), $\Delta$ML(HF) results exhibit
a stronger seed-based spread, illustrated by the broader yellow area.
However, $\Delta$ML(HF) still exhibits a relatively good qualitative agreement. The seed-based spread increases (thickening of the yellow area) at points where the model underestimates the HF reference data.  This can be used to estimate the model's reliability; the larger the seed-based spread, the less confident the prediction.
In cases where $\Delta$ML(HF) predictions deviate significantly from
the base model ML(DFT$_{\rm{E,F}}$), the seed-based spread increases,
this is especially the case for optical modes.
These discrepancies can be largely attributed to the potential energy surface of HF compared to DFT, particularly at the $\Gamma$-point, where it exhibits a stronger curvature. This is not well captured in the $\Delta$-learning process, as it excludes forces at the HF level.

Having verified that our $\Delta$-learning approach can be used to predict HF phonon 
dispersions, we now turn to post-HF methods.

Fig.~\ref{fig:cphon} d) shows phonon dispersions at the level of $\Delta$ML(MP2) (gold area), $\Delta$ML(CCSD) (red dashed area), and $\Delta$ML(CCSD(T)) (purple dotted area).
We first notice that the seed-based spread of these methods is smaller than for
$\Delta$ML(HF). This is attributed to the fact that these methods yield phonon 
frequencies closer to the ones obtained with DFT.
We further corroborate this by observing that $\Delta$ML(MP2), the result closest to DFT, exhibits lowest seed-based spread, in particular at the L point at around 1000$\,$cm$^{-1}$ (which also exhibits the largest seed-based spread overall).

The main differences between $\Delta$ML(MP2), $\Delta$ML(CCSD) and $\Delta$ML(CCSD(T))
are observed for the optical modes at high frequencies.
$\Delta$ML(CCSD) yields the highest frequencies for these modes and $\Delta$ML(CCSD(T))
is between DFT and $\Delta$ML(CCSD).
Compared to experimental findings obtained using neutron scattering~\cite{cexp}  (see $\Gamma$-X)
and Raman spectroscopy~\cite{cgammaexp} (shown at $\Gamma$), $\Delta$ML(CCSD(T))  yields the most 
accurate results.
This confirms expectations based on findings in molecular quantum 
chemistry~\cite{Bartlett2007}, where CCSD(T)
also predicts vibrational frequencies with high accuracy.
However, note that 200 single point CCSD(T) calculations for the studied
supercell consume about 750.000 cpuhs.

For further comparison, Fig.~\ref{fig:cphon} d) depicts QMC data at 
$\Gamma$ (black plus)~\cite{cqmcgamma}, which is also considered an accurate benchmark method, and agrees with our $\Delta$ML(CCSD) results.

We note again that the highest acoustic modes at the L point features particularly large seed-based spread, indicating that these results are less reliable due to the limited supercell size. 

Section 1.3 of the SI~\cite{si} shows all individual seed results explicitly.

We have also investigated the dependence of the results
discussed above on the size of the training data set.
The corresponding phonon dispersions created by MLFFs trained only on around half the training data (98 points) can be found in Sec. 1.1 of the SI~\cite{si}; they are largely similar, but show stronger seed-based spread (except for ML(DFT$_{\rm{E}}$), where primarily the shape of the broadening changes to include imaginary modes).

\subsection{\label{lih}Lithium Hydride}

\begin{figure*}[t]
  \includegraphics[width=\textwidth]{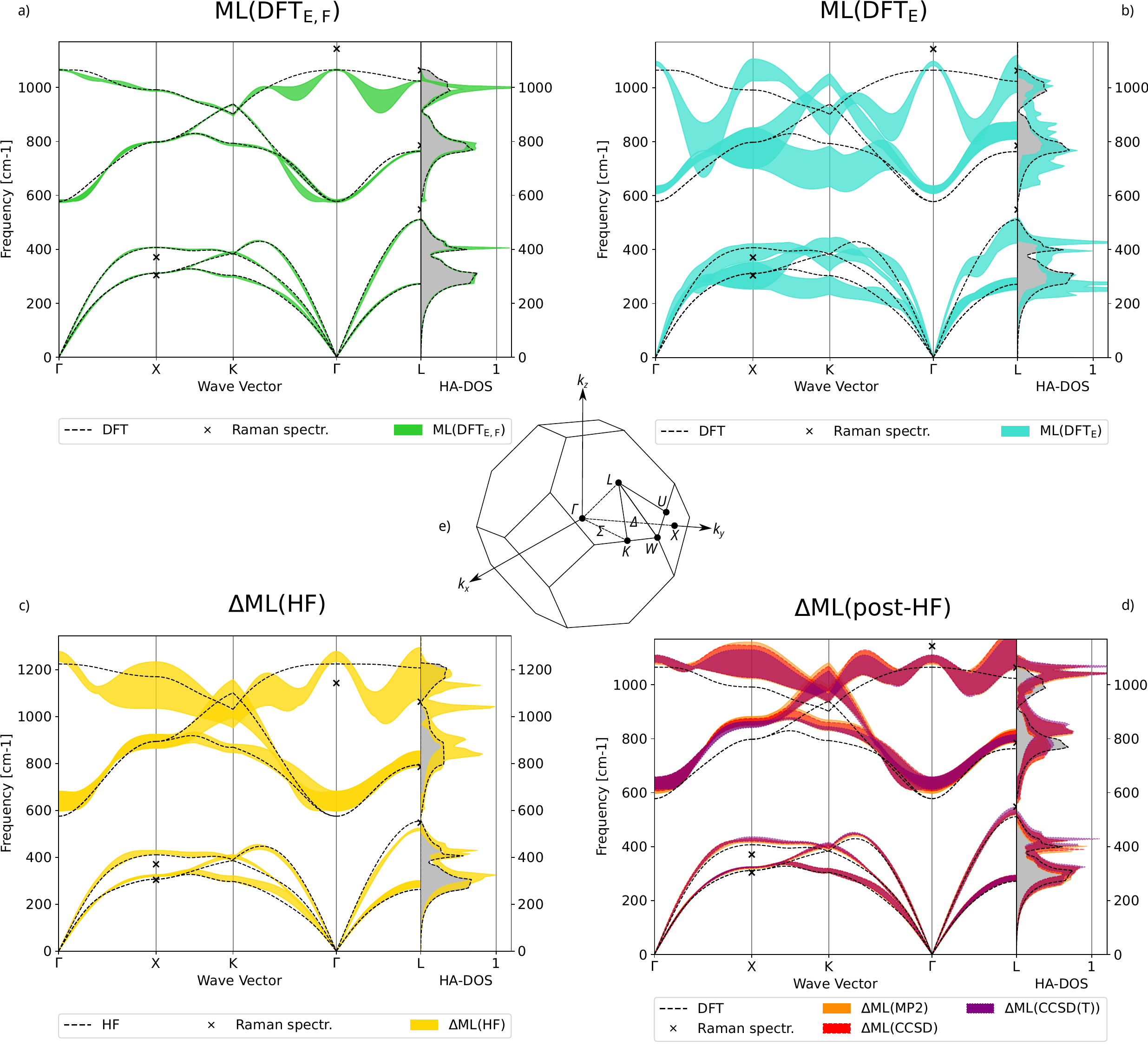}
  \caption{Phonon dispersions and DOS comparisons for LiH done with \texttt{MACE};  Raman spectroscopy data (black x) from~\cite{lihexp}. a) ML(DFT$_{\rm{E,F}}$) across various seeds (green area) vs. DFT results (black dashed line). b) ML(DFT$_{\rm{E}}$) (trained only on E) across various seeds (cyan area) vs. DFT results (black dashed line). c) $\Delta$ML(HF) across various seeds (yellow area) vs. HF results (black dashed line). d) Different $\Delta$ML(WFT) results across various seeds (gold, red dashed, purple dotted areas) vs. DFT results (black dashed line). e) Brillouin Zone with relevant high symmetry points for cubic LiH.}
  \label{fig:lihphon}
\end{figure*}

\begin{figure*}[t]
  \includegraphics[width=\textwidth]{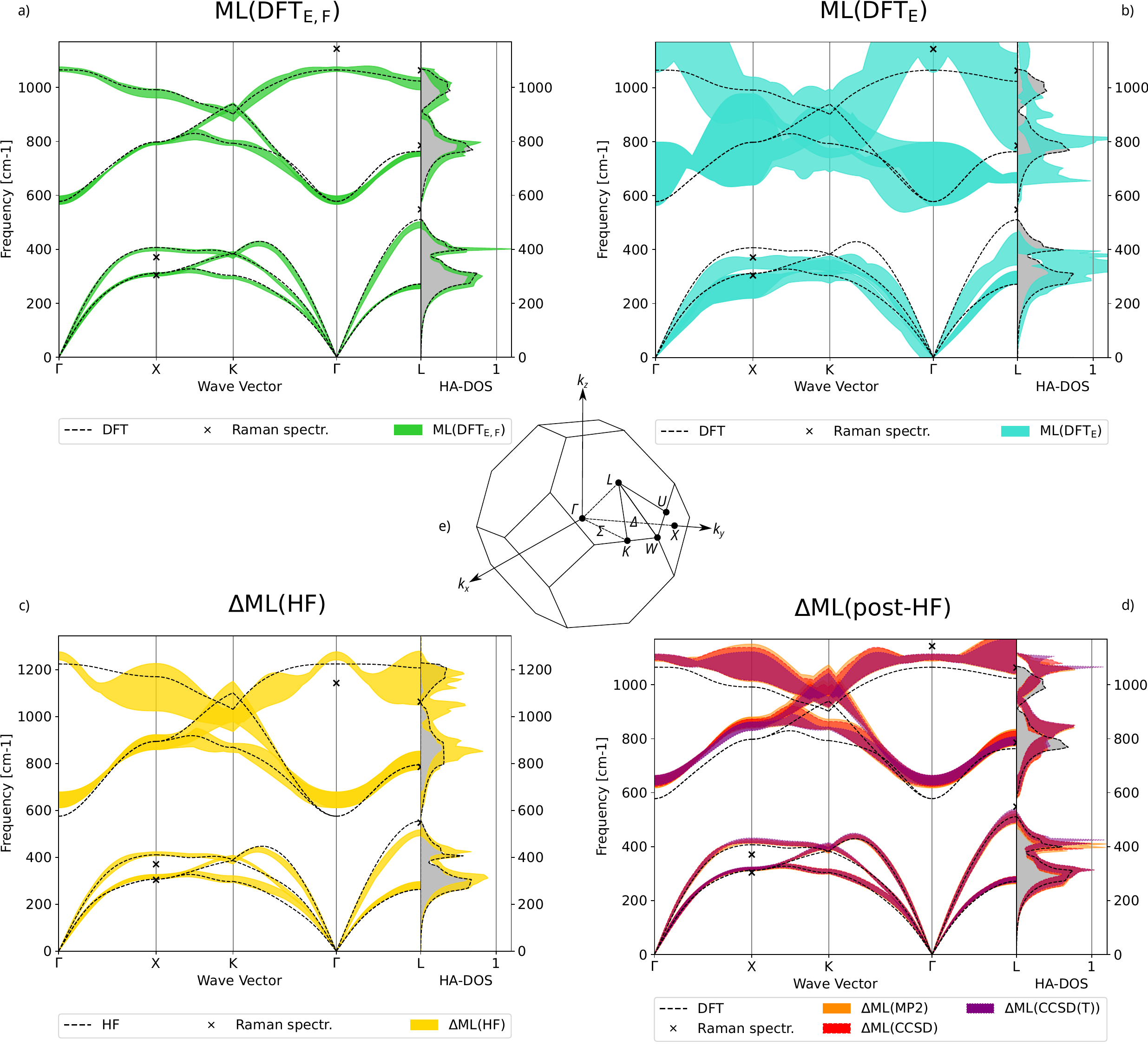}
  \caption{Phonon dispersions and DOS comparisons for LiH done with \texttt{QNEP} and $\Delta$\texttt{MACE};  Raman spectroscopy data (black x) from~\cite{lihexp}. a) ML(DFT$_{\rm{E,F}}$) across various seeds (green area) vs. DFT results (black dashed line). b) ML(DFT$_{\rm{E}}$) (trained only on E) across various seeds (cyan area) vs. DFT results (black dashed line). c) $\Delta$ML(HF) across various seeds (yellow area) vs. HF results (black dashed line). d) Different $\Delta$ML(WFT) results across various seeds (gold, red dashed, purple dotted areas) vs. DFT results (black dashed line). e) Brillouin Zone with relevant high symmetry points for cubic LiH.}
  \label{fig:lihphonqnep}
\end{figure*}

We now turn to the discussion of results obtained for the lithium hydride (LiH) solid.
In this case, we use the local density approximation (LDA) for the exchange and correlation density functional,
which was shown to agree better with experiment for lattice
dynamic properties compared to DFT-PBE~\cite{BARRERA2005119}.
The training data, including structures, energies and forces, for ML(DFT$_{\rm{E,F}}$) were taken from an MD trajectory with $2000$ steps.
More details are described in Sec.~\ref{lihcomp}.
This number of configurations was necessary to achieve acceptable dispersions.

Fig.~\ref{fig:lihphon} a) shows that the phonon dispersion computed with ML(DFT$_{\rm{E,F}}$) (green area) is mostly
in good agreement with results generated by DFT (black dashed line). The acoustic modes show almost perfect agreement
for all seeds, whereas the optical modes exhibit regions with significant discrepancies, especially when
approaching the $\Gamma$ point from K and L. These will be addressed in Fig.~\ref{fig:lihphonqnep}. For a depiction of each individual seed, see Sec. 2.3 of the SI~\cite{si}.

Phonon frequencies at the $\Gamma$ point are all in
relatively good agreement, but it should be noted that this is partly a consequence
of the fact that the splitting of the LO-TO (longitudinal and transversal optical) mode at this 
point is  determined from a non-analytic term correction (NAC) that 
is computed employing Born-effective charges at the level of 
DFT~\cite{Gonze1997}.
This correction is identical for ML(DFT$_{\rm{E,F}}$) and DFT 
phonon dispersions, explaining the perfect agreement of the 
splitting at $\Gamma$.

Fig.~\ref{fig:lihphon} b) shows phonon dispersions generated by ML(DFT$_{\rm{E}}$)
(cyan area), which was trained using 200 configurations only

 as this is the number of systems that are feasible using periodic CCSD(T) 
calculations without significant computational cost.
Note that 200 single point CCSD(T) calculations for the the studied
supercell consume about 10.000 cpuhs.
Compared to DFT (black dashed line), ML(DFT$_{\rm{E}}$) exhibits significant discrepancies especially for the optical modes.
The only exception is again the $\Gamma$ point due to inclusion of the NAC.

Following the same procedure as in the previous section, 
we then apply the $\Delta$-learning approach to HF and compare with the reference HF phonon dispersion (black dashed line),
depicted in Fig.~\ref{fig:lihphon} c) (yellow area). The acoustic modes overlap quite well with just a slightly larger
spread than ML(DFT$_{\rm{E,F}}$).
However, the agreement of the optical modes is much worse, although better than in the case of ML(DFT$_{\rm{E}}$).

Also notable is very large seed-based spread at the highest optical mode at the X point.

In Fig.~\ref{fig:lihphon} d), the post-HF $\Delta$-learning approaches are plotted for
MP2, CCSD and CCSD(T) (gold, red dashed, purple dotted areas).
Interestingly, all post-HF methods yield phonon dispersions that agree with each other relatively well for LiH.
Note that the DFT-based NAC was also used in this case, which is slightly less justified for these methods.
However, we stress that this correction only affects the splitting of the LO and TO modes at $\Gamma$.
Since this splitting agrees relatively well between DFT and all post-HF methods at the other k-points,
this approximation seems well justified, and finds some more legitimization in Sec.~\ref{vacf}.
The capability to compute Born effective charges for the post-HF methods will be the topic of future work.

Overall, we observe little seed-based spread in the acoustic modes, though slightly larger compared to ML(DFT$_{\rm{E,F}}$).
The seed-based spread again becomes very large for the optical modes.
The largest spread can be seen in $\Delta$ML(MP2), with $\Delta$ML(CCSD(T)) having the smallest.
We observe again a particularly large seed-based spread at the X point.
It is noteworthy that the frequencies of the optical modes are larger in all post-HF methods compared to DFT.

As before, the phonon dispersions generated by MLFFs trained on around half the training data (998 for ML(DFT$_{\rm{E,F}}$), 85 for the $\Delta$ML(WFT)) reveal broadly speaking similar performance, with stronger seed-based spread especially for $\Delta$-learning results (see Sec. 2.1 of the SI~\cite{si}).

An attempt to mitigate the oscillatory behavior in the dispersion of
the optical frequencies is found in Fig.~\ref{fig:lihphonqnep}. Here, we used \texttt{QNEP} instead of \texttt{MACE} for the ML(DFT$_{\rm{E,F}}$) and ML(DFT$_{\rm{E}}$) results, the former also applied as the base for the $\Delta$-learning results. The NN trained through \texttt{QNEP} includes long-range electrostatic interaction, reducing the oscillations,
but increasing error and seed-based spread at the upper optical modes of the chosen high symmetry points.

None of the LiH phonon results - whether ML, DFT or HF - reproduce the  experimental Raman spectroscopy results~\cite{lihexp} particularly well. We will investigate how strong a role the anharmonicity of LiH~\cite{lihanh} plays here in Sec.~\ref{vacf}.

\subsection{Velocity Autocorrelation Function}\label{vacf}

In addition to the phonon dispersions of diamond and LiH, Figs.~\ref{fig:cphon} and \ref{fig:lihphon} show the vibrational densities of states in the harmonic approximation (HA-DOS).
The MLFFs can also be used to estimate the density of states including contributions beyond the harmonic approximation by performing long MD simulations and computing the
Fast Fourier Transform (FFT) of the Velocity Autocorrelation Function (VACF) from the obtained trajectories. 
The vibrational density of states computed from the VACF is denoted as VACF-DOS.

To expedite this calculation process, only two of the five seeds from before are used to create the shown area/spread for both HA-DOS and VACF-DOS.

\begin{center}
 \begin{figure}[hbt!]
   \includegraphics[width=\textwidth]{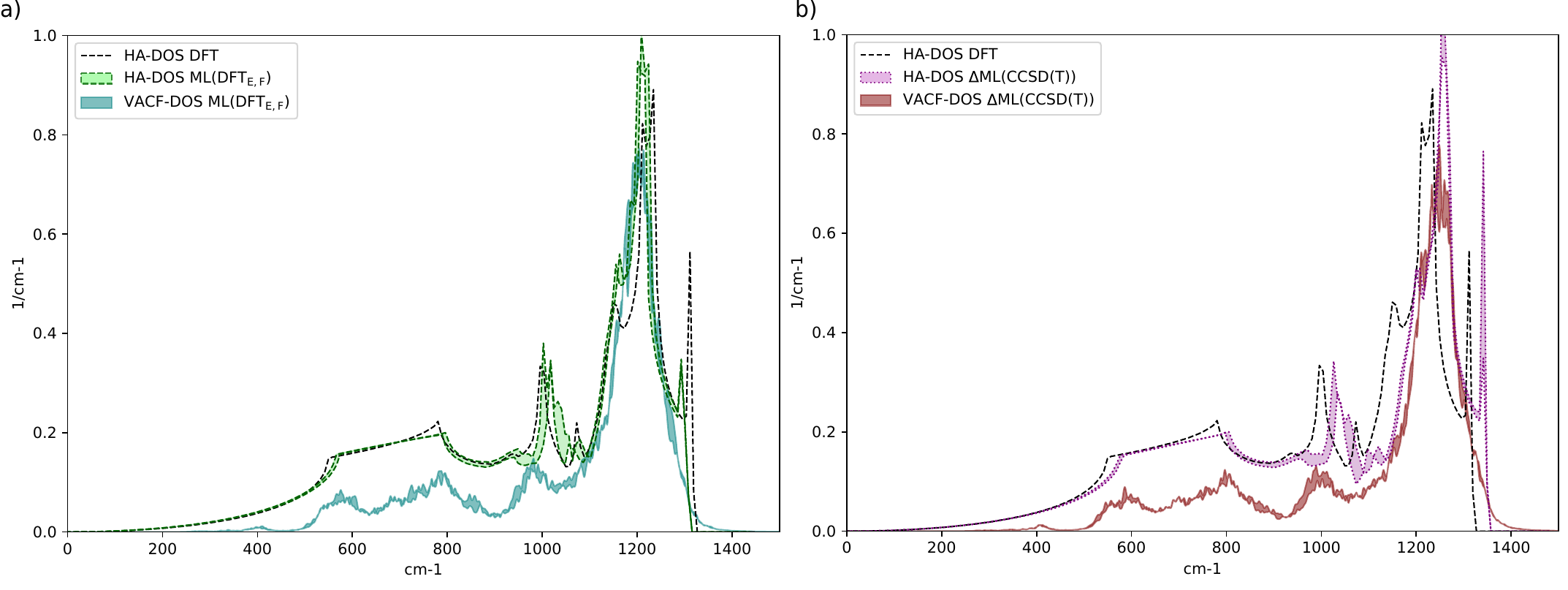}
   \caption{HA- and VACF-DOS comparisons for diamond. a) Comparison between DFT HA-DOS (black dashed line), ML(DFT$_{\rm{E,F}}$) HA-DOS (green dashed area) and ML(DFT$_{\rm{E,F}}$) VACF-DOS (teal area). b) Comparison between DFT VDOS (black dashed line), $\Delta$ML(CCSD(T)) HA-DOS (purple dotted area) and $\Delta$ML(CCSD(T)) VACF-DOS (brown area). HA-DOS calculated as explained in Sec.~\ref{sec:comp}, using either ML or DFT methods; areas created using two of the five seeds from Sec.~\ref{c}. VACF generated using ML MDs at $300$ K. }
   \label{fig:cvacf}
 \end{figure}
\end{center}

We first apply this method to diamond. Fig.~\ref{fig:cvacf} depicts the HA-DOSs as generated by DFT (black dashed line), ML(DFT$_{\rm{E,F}}$) (green dashed area, Fig.~\ref{fig:cvacf} a)) and $\Delta$ML(CCSD(T)) (purple dotted area, Fig.~\ref{fig:cvacf} b)).
As one could already observe in Fig.~\ref{fig:cphon} d), these results are mostly similar, except for the optical modes, where $\Delta$ML(CCSD(T)) is shifted to higher
frequencies compared to ML(DFT$_{\rm{E,F}}$).
To obtain the DOS including effects beyond the HA, we performed 
a molecular dynamics simulation (MD) at  $300\,$K once using ML(DFT$_{\rm{E,F}}$) (green area, Fig.~\ref{fig:cvacf} a)) and once using $\Delta$ML(CCSD(T)) (brown area, Fig.~\ref{fig:cvacf} b)).
The VACF-DOSs were obtained by a FFT of the VACFs.
However, the peak positions in both VACF-DOSs are similar to those 
in the respective HA-DOS. Remaining discrepancies in peak 
height and shape are owed to the finite mesh and supercell sizes 
used in the VACF calculations.
Although MLFFs can be used to generate MD trajectories
efficiently, it is still impossible to perform calculations of
VACF-DOSs using supercells that correspond to the dense meshes
used for the HA-DOSs.

\begin{center}
 \begin{figure}[hbt!]
   \includegraphics[width=\textwidth]{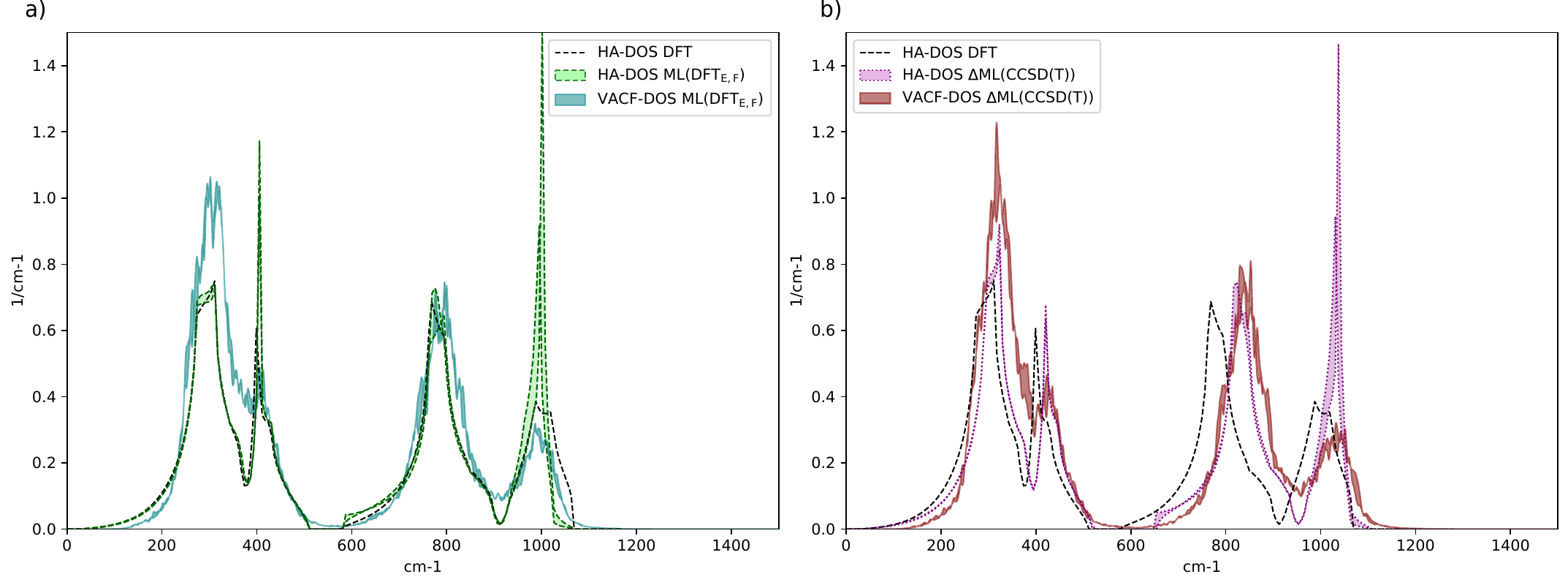}
   \caption{HA- and VACF-DOS comparisons for LiH. a) Comparison between DFT HA-DOS (black dashed line), ML(DFT$_{\rm{E,F}}$) HA-DOS (green dashed area) and ML(DFT$_{\rm{E,F}}$) VACF-DOS (teal area). b) Comparison between DFT HA-DOS (black dashed line), $\Delta$ML(CCSD(T)) HA-DOS (purple dotted area) and $\Delta$ML(CCSD(T)) VACF-DOS (brown area). HA-DOS calculated as explained in Sec.~\ref{sec:comp}, using either ML or DFT methods; areas created using two of the five seeds from Sec.~\ref{lih}. VACF generated using ML MDs at $20$ K. }
   \label{fig:lihvacf}
 \end{figure}
\end{center}

\begin{center}
 \begin{figure}[hbt!]
   \includegraphics[width=\textwidth]{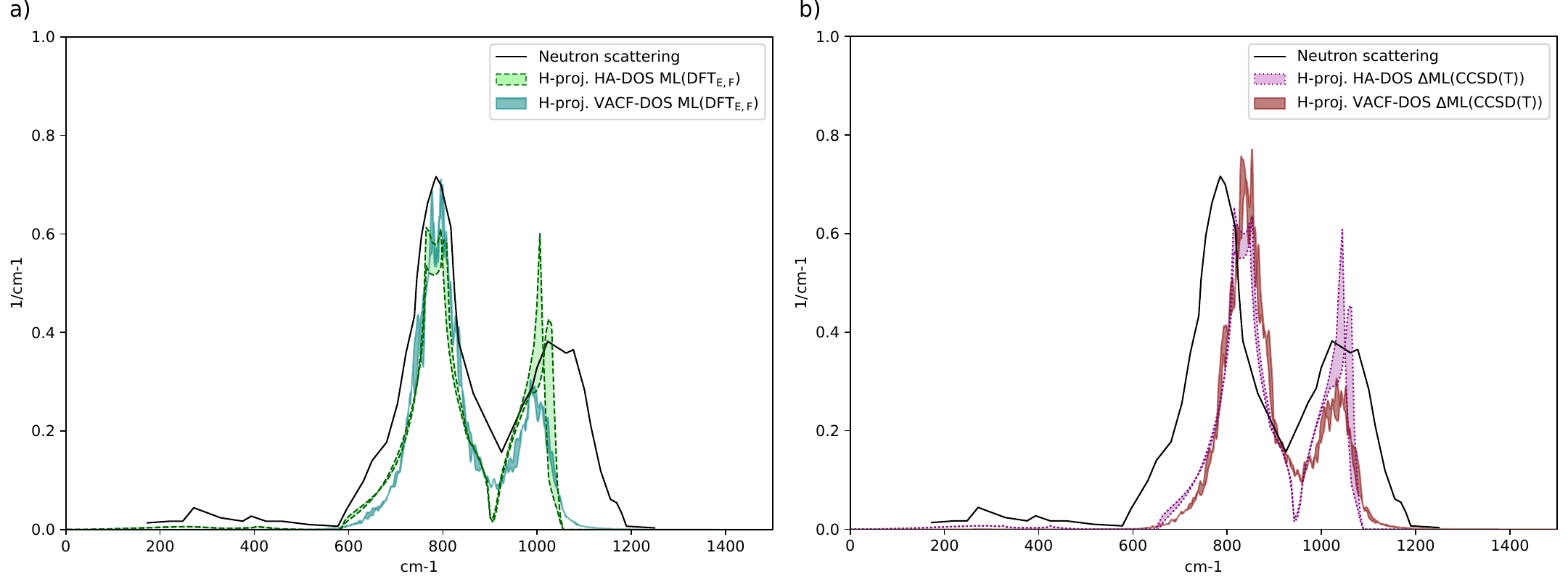}
   \caption{H-projected HA- and VACF-DOS comparisons for LiH. a) Comparison between H-projected neutron scattering DOS~\cite{lihdosexp} (black line, scaled up for better visibility), H-projected ML(DFT$_{\rm{E,F}}$) HA-DOS (green dashed area), and H-ptojected ML(DFT$_{\rm{E,F}}$) VACF-DOS (teal area). a) Comparison between H-projected neutron scattering DOS~\cite{lihdosexp} (black line, scaled up for better visibility), H-projected $\Delta$ML(CCSD(T)) HA-DOS (purple dotted area), and H-projected $\Delta$ML(CCSD(T)) VACF-DOS (brown area). HA-DOS calculated as explained in Sec.~\ref{sec:comp}, using ML methods. Spreads given by two different seed results. VACF generated using ML MDs at $20$ K.}
   \label{fig:lihvacfexp}
 \end{figure}
\end{center}

Next, we turn to LiH to see whether anharmonic effects have a larger contribution to the vibrational properties. 
In Fig.~\ref{fig:lihvacf}, the same procedure as before is plotted for LiH. The HA-DOS generated by DFT (black dashed line), ML(DFT$_{\rm{E,F}}$) (green dashed area, Fig.~\ref{fig:lihvacf} a)) and $\Delta$ML(CCSD(T)) (purple dotted area, Fig.~\ref{fig:lihvacf} b)) largely overlap, with $\Delta$ML(CCSD(T)) exhibiting a shift to higher frequencies in the optical modes (as seen before in Fig.~\ref{fig:lihphon} d)). The VACF-DOSs are once again calculated from an MD run using ML(DFT$_{\rm{E,F}}$) (green area, Fig.~\ref{fig:lihvacf} a)) and $\Delta$ML(CCSD(T)) (purple area, Fig.~\ref{fig:lihvacf} b)), this time at $20$ K to match with the neutron scattering experiment shown in Fig.~\ref{fig:lihvacfexp}.
Note that this temperature should be sufficient to excite all vibrations in the studied frequency range. 

However, the VACF-DOSs and HA-DOSs once again largely overlap, showing that direct anharmonic effects do not have a strong effect on the DOS. Looking at the peak originating from the optical modes at high frequencies in Fig.~\ref{fig:lihvacf} b), this also legitimizes the use of the DFT Born charges for the $\Delta$ML(CCSD(T)) calculations: The VACF-DOS is Born-charge independent, yet the two overlap at this peak (the only one affected by the non-analytical term correction).

In Fig.~\ref{fig:lihvacfexp}, we show a comparison between experimental neutron scattering data from Ref.~\cite{lihdosexp} (black line, scaled up for better visibility) and the
H-projected (HA/VACF-)DOSs from Fig.~\ref{fig:lihvacf}.

We find that the experimental peak around $800\,$cm$^{-1}$ matches well with the ML(DFT$_{\rm{E,F}}$) results, the experimental peak around $1100$ cm$^{-1}$ is in better agreement with the corresponding peaks from $\Delta$ML(CCSD(T)).

Remaining errors may stem from the experiments, mismatches in lattice constants between experiments and training data, and finite size effects.
Additionally, nuclear quantum effects may also be significant.

\end{multicols}

\clearpage

\begin{multicols}{2}

\section{\label{disc}Discussion}

In this work we have used $\Delta$-learning to obtain MLFFs for periodic solids at
MP2, CCSD and CCSD(T) levels of accuracy.
An advantage of the employed $\Delta$-learning approach is that it
circumvents the need for calculating atomic forces at the desired level of
quantum chemical many-electron theories, such as MP2, CCSD or CCSD(T) theory.
Instead, the presented approach combines a more precise MLFF trained on
DFT energies and forces, with an MLFF trained on MP2, CCSD or CCSD(T)
energies only.
We have assessed the precision of the ($\Delta$-learning) MLFFs by comparing to (HF) DFT 
reference results using different seeds for \texttt{MACE}.
Throughout this procedure, identical training data structures
were used.
The accuracy of the MLFFs at the level of DFT and CCSD(T) theory
was assessed by comparing to experimental vibrational frequencies.
We have also demonstrated that training MLFFs on total energies alone,
as would be desired for methods where atomic forces are not already
implemented, does not yield very precise force fields.
Reducing the training data set size 
does not lead to significant deterioration of results, as shown in Sec. 1.1 and 2.1 of the SI~\cite{si}.

In the case of carbon diamond, we found that the acoustic phonon dispersions agree
very well with experimental findings at both the DFT-PBE and CCSD(T) levels of theory.
ML(DFT$_{\rm{E,F}}$) is capable of reproducing the DFT results with high precision.
The seed-based spread of $\Delta$ML is sufficiently small to differentiate the computed phonon frequencies between DFT-PBE,
MP2, CCSD and CCSD(T).
For the optical modes, we find that CCSD(T) theory yields higher frequencies in better agreement with experiment compared to DFT-PBE.

Our findings for LiH are partly similar to those of diamond.
For the optical modes, CCSD(T) theory yields higher frequencies
than DFT-PBE.
Moreover, the good agreement between MP2, CCSD and CCSD(T) theory
indicates that the employed level of electronic structure
theory yields highly reliable potential energy surfaces.
In contrast to the results for diamond,
we find for LiH that long-range electrostatic effects play an important role in the training of precise MLFFs regardless of the level of theory.
In particular, we found that disregarding long-range effects leads to
unphysical oscillations in the phonon frequency dispersion of the 
optical modes. We have mitigated this behaviour using \texttt{QNEP}
at the level of ML(DFT$_{\rm{E,F}}$).

Using ML(DFT$_{\rm{E,F}}$) and $\Delta$ML(CCSD(T)), it is possible to perform
MD simulations for relatively long time scales needed to compute
converged VACFs ($30000$ steps for cells containing $512$ atoms, see Fig.~\ref{fig:workflow} and Table~\ref{tab:settings}).
Our calculations show that anharmonic effects yield only
small changes in the VDOS.
However, both DFT-LDA and CCSD(T) phonon dispersions as well as
vibrational densities of states
in the harmonic approximation do not match the experimental findings well.
Investigating H-projected vibrational densities that can be directly compared to neutron scattering yields better results at the DFT-LDA level for lower optical modes, while for the higher optical modes the CCSD(T) level fares best.

The remaining discrepancies between theory and experiment
cannot be fully explained and may necessitate inclusion  
of effects from lattice expansion and perhaps non-adiabaticity.
Additionally, nuclear quantum effects could also play a significant
role. This requires further investigation beyond the scope of the present work. However, The published training data and the outlined $\Delta$ML
approach can form the basis of future work employing path integral molecular dynamics

In all simulations, the computational bottleneck remains the generation
of training data at the higher levels of theory. In particular,
the sizes of the simulation supercells are the main limitation.
In the case of asymmetric supercells,
e.g. carbon diamond, we find that this leads to lower precision of the respective
MLFFs at certain $k$-points.

Taken together, our findings indicate that improving MLFFs by incorporating forces also at higher level of theory is particularly beneficial compared to relying solely on the availability of single-point energies, even when the latter can be computed using larger simulation cells. We therefore conclude that implementations including forces are highly likely to offer substantial advantages despite the associated increase in computational cost. Consequently, we are pursuing the implementation of forces within periodic CCSD(T), which is currently underway. In addition, our results suggest that investigating architectures that explicitly account for long-range interactions represents a particularly promising direction for future work.

\section{\label{meth}Methods}

This section summarizes details of all employed methods and
is organized as follows.
Sec.~\ref{sec:workflow} explains the general workflow
used to obtain the vibrational density of states
and phonon dispersions for the materials studied in this work.
Sec.~\ref{sec:delta} describes the $\Delta$-learning approach needed
to compute vibrational properties at the level of HF, MP2, CCSD and CCSD(T) theory.
Sec.~\ref{sec:comp} summarizes details of the electronic structure
theory calculations used to generate the training data as well as
settings used to train the MLFFs. 

\subsection{\texorpdfstring{$\Delta$-learning}{Delta-Learning}}\label{sec:delta}

We first explain additional details of the $\Delta$-learning approach
used to obtain MLFFs at the level of wavefunction theories.
Here, wavefunction theory (WFT) stands for Hartree-Fock (HF),
second-order M\o ller-Plesset perturbation (MP2),
coupled-cluster singles and doubles (CCSD) or
coupled-cluster singles, doubles plus perturbative triples (CCSD(T)) theory.
The $\Delta$-learning approach is necessary because the implementation
of periodic MP2 and CC theory used in the present work does not yet provide atomic forces.

MLFFs trained on total energies alone do not achieve the level of precision needed for reliable
phonon dispersions and VDOSs, as shown in Figs.~\ref{fig:cphon}, \ref{fig:lihphon}, \ref{fig:lihphonqnep} b).
To overcome this limitation, the $\Delta$-learning approach combines training on DFT
energies and forces with training on differences between energies obtained at the
levels of DFT and WFT.
We use a $\Delta$-learning approach, which involves training two MLFFs:

\begin{enumerate}[\hspace{0.9cm}] 
    \item ML(DFT$_{\rm{E,F}}$) refers to an MLFF trained on DFT energies (E) and atomic forces (F) and
    \vspace{0.25cm}
    \item ML(WFT-DFT) refers to an MLFF trained on the difference in E between structures calculated by DFT and the relevant WFT-based method, $\Delta E = E_{\rm{WFT}} - E_{\rm{DFT}}$.
\end{enumerate}

We employ \texttt{MACE} or \texttt{QNEP} to train ML(DFT$_{\rm{E,F}}$) by providing \texttt{xyz}-files containing
the structural information for a set of supercells with corresponding total energies and atomic forces.
These structures are taken from an MD trajectory. System specific details
of the MD simulations are summarized in Table~\ref{tab:settings}.

More details about the settings used by \texttt{MACE} and \texttt{QNEP} during training are specified in Sec.~\ref{sec:mace}.

ML(WFT-DFT) is trained using \texttt{MACE} by providing corresponding \texttt{xyz}-files
that contain the structural information along with the corresponding total energy differences,
$\Delta E = E_{\rm{WFT}} - E_{\rm{DFT}}$.

To predict energies at the level of WFTs
for a given structure,
we simply add the predicted energies of
ML(DFT$_{\rm{E,F}}$) and ML(WFT-DFT):
\begin{multline}
E_{\Delta \mathrm{ML(WFT)}} = E_{\mathrm{ML(DFT_{\rm{E,F}})}} \\
+ E_{\mathrm{ML(WFT-DFT)}}.
\end{multline}
The atomic forces at the level of WFTs
are also obtained by adding the
predicted atomic forces of both MLFFs.
Note that, as a shorthand, we always refer to the total result as $\Delta$ML(WFT) (see the black dashed arrow branch in Fig.~\ref{fig:workflow}): It is the result of adding ML(WFT-DFT) (sheer yellow-red gradient rounded rectangle) to ML(DFT$_{\rm{E,F}}$) (green rounded rectangle).

In this work, we train $\Delta$ML(WFT) models for HF, MP2,
CCSD and CCSD(T). $\Delta$ML(HF) also serves as a ``sanity check", as we can produce HF-level phonon dispersions using \texttt{VASP} without ML, and compare to $\Delta$ML(HF) (see Fig.~\ref{fig:cphon}, \ref{fig:lihphon}, \ref{fig:lihphonqnep} c)). The other $\Delta$-learning results are found in Fig.~\ref{fig:cphon}, \ref{fig:lihphon}, \ref{fig:lihphonqnep} d).

\subsection{Workflow}\label{sec:workflow}

 \begin{figure*}
 \begin{minipage}{0.6\textwidth}
 \hspace{0.5cm}
   \includegraphics[width=0.9\textwidth]{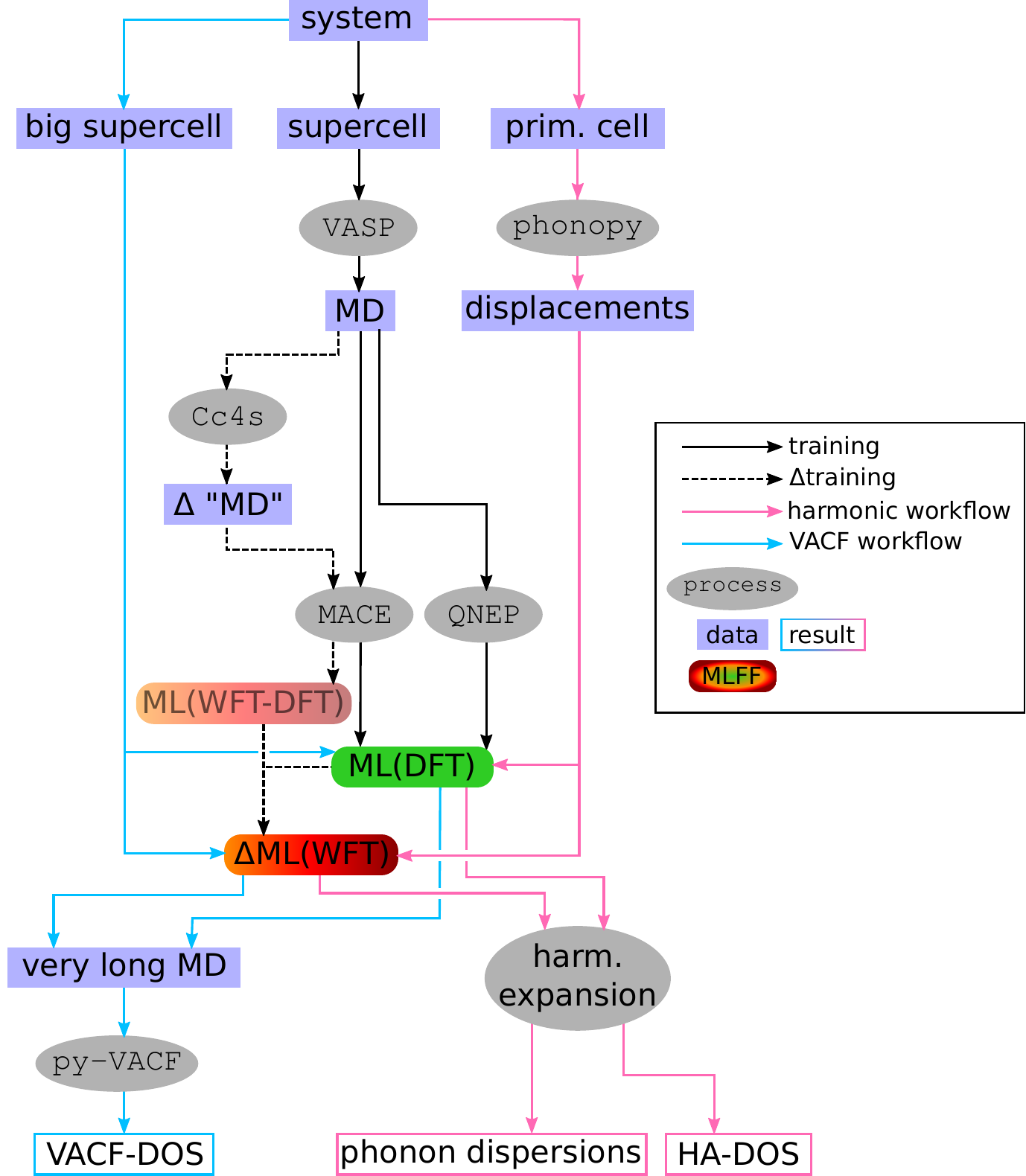}
   \end{minipage}
  \begin{minipage}{0.35\textwidth}
   \caption{Workflow for this paper. First, a system (diamond or LiH) is chosen. In the black branch, a supercell of the conventional cell is created (for exact settings always see Table~\ref{tab:settings}) using \texttt{VASPKIT}\cite{VASPKIT}. From this initial configuration, \texttt{VASP}\cite{vasp1,vasp2,vasp3} is used to create an MD. \texttt{MACE}\cite{Batatia2022mace,Batatia2022Design} or \texttt{QNEP}\cite{qnep1,qnep2,qnep3} are trained on said MD, yielding ML(DFT).
   In the black dashed branch, 200 steps of the previous MD are then fed into \texttt{Cc4s}\cite{cc4s_website}, which generates WFT level energies for each configuration (and forces in the case of HF). The energy \emph{differences} from this $\Delta$ ``MD" is again used to train \texttt{MACE}, yielding ML(WFT-DFT); predictions from ML(WFT-DFT) added to ML(DFT$_{\rm{E,F}}$) creates the results from $\Delta$ML(WFT).
   In the pink branch, a relaxed primitive cell is fed into \texttt{phonopy}\cite{phonopy-phono3py-JPCM,phonopy-phono3py-JPSJ}, yielding displacement configurations. For these, the MLFFs predict energies and forces, which are then interpreted by \texttt{phonopy} for a harmonic expansion to generate the phonon dispersions and HA-DOS.
   In the light blue VACF branch, the ML(DFT$_{\rm{E,F}}$) and $\Delta$ML(CCSD(T)) are both used to create two very long MDs which are handed to \texttt{py-VACF}\cite{vacf} to yield the VACF-DOS.}

   \label{fig:workflow}
   \end{minipage}
 \end{figure*}

A workflow of the data and software used is depicted in Fig.~\ref{fig:workflow}. In addition, a summary of the most relevant settings is given in Table~\ref{tab:settings}, containing information about the employed
cell sizes and computational parameters.

\definecolor{b}{HTML}{b2b2ff}
\definecolor{pro}{HTML}{b2b2b2}
\definecolor{ml}{HTML}{00ff00}
\newcommand{\mlbox}[1]{
  \tikz[baseline=(text.base)]{
    \node[
      inner sep=2pt,
      text=black,
      align=center,
      rectangle,
      shade,
      left color=ml,
      right color=ml,
      rounded corners=2pt
    ] (text) {#1};
  }
}
\definecolor{gr1}{HTML}{ff8c00}
\definecolor{gr2}{HTML}{ff0000}
\definecolor{gr3}{HTML}{ff0000} 
\newcommand{\gradientbox}[1]{
  \tikz[baseline=(text.base)]{
    \node[
      inner sep=2pt,
      text=black,
      align=center,
      rectangle,
      shade,
      left color=gr1,
      middle color=gr2,
      right color=gr3,
      rounded corners=2pt
    ] (text) {#1};
  }
}

\newcommand{\opbox}[1]{
  \tikz[baseline=(text.base)]{
    \node[
      inner sep=2pt,
      text=black,
      align=center,
      rectangle,
      shade,
      left color=gr1,
      middle color=gr2,
      right color=gr3,
      opacity=0.5,
      rounded corners=2pt
    ] (text) {#1};
  }
}

\definecolor{ha}{HTML}{ff69b4}
\definecolor{va}{HTML}{00bfff}
\newcommand{\coloredborderbox}[2]{
  \tikz[baseline=(text.base)]{
    \node[
      draw=#1,          
      line width=1pt,
      inner sep=4pt,
      text=black,
      rectangle,
      rounded corners=2pt,
      fill=none
    ] (text) {#2};
  }
}

The black arrows in Fig.~\ref{fig:workflow} depict the flow of data for training MLFF:
After choosing a \textbf{system} (diamond or LiH),
a selected \textbf{supercell} is generated using, in our case, \texttt{VASPKIT}~\cite{VASPKIT} (for details see Table~\ref{tab:settings}) and used to perform
\textbf{MD} simulations at the level of DFT-PBE (diamond) or DFT-LDA (LiH) using \texttt{VASP}~\cite{vasp1,vasp2,vasp3}. The MD trajectory including computed
energies and forces is fed into \texttt{MACE}~\cite{Batatia2022mace,Batatia2022Design} (or, in the case of Fig.~\ref{fig:lihphonqnep}, \texttt{QNEP}~\cite{qnep1,qnep2,qnep3}) for training and, still following the black arrows, produces \textbf{ML(DFT)}. ML(DFT$_{\rm{E,F}}$) and ML(DFT$_{\rm{E}}$) refers to training with \texttt{MACE} or \texttt{QNEP}
on both energies and forces or energies alone, respectively.

Black dashed arrows in Fig.~\ref{fig:workflow} illustrate the $\Delta$-learning 
approach used in this work.
We employ a subset of structures from the DFT MD to compute MP2, CCSD and CCSD(T)
energies. The CC calculations are performed using \texttt{Cc4s}~\cite{cc4s_website}.
The previously generated DFT energies for the same structures are subtracted to yield \textbf{$\Delta$``MD"} training data, which is used in \texttt{MACE} to train \textbf{ML(WFT-DFT)}. WFT stands for HF, MP2, CCSD or CCSD(T).
\textbf{$\Delta$ML(WFT)} is a MLFF, denoting predictions from \textbf{ML(WFT-DFT)} and \textbf{ML(DFT$_{\rm{E,F}}$)} added together.
For further details on the $\Delta$-learning process see Sec.~\ref{sec:delta}. 

The workflow used to compute phonon dispersions and VDOSs in the harmonic
approximation is outlined by the pink arrows. First, a relaxed
\textbf{primitive cell} of the relevant system is fed into \texttt{phonopy}~\cite{phonopy-phono3py-JPCM,phonopy-phono3py-JPSJ}, which yields structures for
different \textbf{displacements}.
For these, the MLFFs predict energies and forces, which are fed back into \texttt{phonopy} to compute the \textbf{phonon dispersions} and \textbf{HA-DOS} shown in Figs.~\ref{fig:cphon} (diamond) and \ref{fig:lihphon} (LiH) (pink boxes in Fig.~\ref{fig:workflow}).
In the case of LiH we also include the NAC (not shown in Fig.~\ref{fig:workflow}).

In the VACF-DOS workflow (blue arrows), a \textbf{big supercell} (exact settings can be found in Table~\ref{tab:settings}) is used as the initial configuration for a \textbf{very long MD} generated using the MLFFs. The atomic velocities of the very long MD
are used to compute auto-correlation functions, which are Fourier transformed by \texttt{py-VACF}~\cite{vacf}, yielding the \textbf{VACF-DOS} found in Figs.~\ref{fig:cvacf}, \ref{fig:lihvacf}, and \ref{fig:lihvacfexp} (blue box in Fig.~\ref{fig:workflow}).

\subsection{\label{sec:comp}Computational Details}

\begin{table*}[t]
\centering
\caption{\label{tab:settings}Settings for different parts of the workflow. Refers back to Fig.~\ref{fig:workflow}. n are the number of MD steps, r\_max defines up to what distance atoms are included in the descriptor, T is temperature.}
\begin{tabular}{ l  c  c }
\midrule
 & C & LiH \\
 \midrule
 \midrule
supercell      & $2\times2\times1$ ($32$) & $2\times2\times2$ ($64$) \\
n [MD, ML(DFT$_{\rm{E,F}}$)]& $200$ & $2000$  \\
n [$\Delta$ ``MD", ML(WFT-DFT), ML(DFT$_{\rm{E}}$)]& $200$ & $200$  \\
r\_max (\texttt{MACE})  &  $3.0$ \AA  &  $9.0$ \AA \\
cutoff (\texttt{QNEP})  &    & $8.0$, $4.0$ \AA \\
big supercell   & $4\times4\times4$ ($512$) & $4\times4\times4$ ($512$) \\
n [very long MD]& $30000$ & $30000$ \\
T [very long MD]& $300$ K & $20$ K  \\
supercell [displacements] & $4\times4\times4$ ($128$) & $4\times4\times4$ ($128$) \\
mesh [harm. expansion] & $24\times24\times24$ & $24\times24\times24$   \\
\bottomrule
\end{tabular}
\vspace{0.25cm}
\end{table*}

\subsubsection{\texttt{MACE}}\label{sec:mace}

The majority of MLFFs presented in this work are based on the implementation of the
message-passing neural network \texttt{MACE }\cite{Batatia2022mace,Batatia2022Design}.
In the training procedure, five randomly chosen seeds ($5192$, $5391$, $5735$, 
$7271$, and $861$) were employed.
The optimization process took 50 epochs.
The MLFFs were used to compute phonon dispersions and HA-DOSs. 
In the case of the more expensive VACF MD needed for VACF-DOSs, only $5735$ and $861$ were used. For a depiction of each phonon dispersion for each model and seed, see Sec.~1.3 (diamond) and 2.3 (LiH) of the SI.
Table~\ref{tab:settings} summarizes the computational parameters; for example,
the supercell size and the radial cutoff parameter (\texttt{r\_max})
within which atoms are considered for a single descriptor. Note that the hyperparameter
\texttt{r\_max} was 3$\,$\AA$\,$and 9$\,$\AA$\,$for diamond and LiH, respectively.
The larger cutoff for LiH is necessary because its ionic bond character substantially
increases long-range interatomic interactions compared to diamond.

To investigate the dependence on the training data set size we also
trained MLFFs with half the training data set size.
``Half'' sets are defined such that they contain only 98 randomly
picked structures for training, whereas the remaining 102 are used for testing. 
These results are found in Sec~1.1 (2.1 for LiH) of the SI~\cite{si}.
For the RMSEs, see also Sec.~1.2 (2.2 for LiH) of the SI~\cite{si}.
Overall, half sets showed around equal agreement compared to full sets
with partially increased seed-based spread. The RMSEs for all sets and methods in
are in the meV/atom range, except ML(DFT$_{\rm{E}}$) for diamond, which is in the
10$\,$meV/atom range.

\subsubsection{\texttt{QNEP}}\label{sec:qnep}

To investigate the role of long-range forces, MLFFs based on \texttt{QNEP}~\cite{qnep1,qnep2,qnep3} were also employed, specifically in Fig.~\ref{fig:lihphonqnep}. Here, too, five random seeds were used, although they cannot be specifically chosen in the \texttt{QNEP} architecture, but are randomly generated every time a new training run is started. Default settings were used, most importantly \texttt{cutoff} $8.0$ and $4.0$ \AA~radial and angular respectively, and \texttt{charge\_mode} 2, referring to using partial charges from the reciprocal space only. 

\subsubsection{Diamond}\label{ccomp}

The training data structures for diamond are obtained as follows: We use \texttt{VASP}~\cite{vasp1,vasp2,vasp3} to perform an MD simulation at the level of DFT-PBE
for a $32$ atom starting structure ($2\times2\times1$ supercell of the conventional basis cell, lattice constant 3.561$\,$\AA) to produce a $2000$ step MD trajectory.

In order to ensure structural variety in our training structures, we applied a temperature sweep from $0$ to $500\,$K employing a Langevin thermostat.

A randomly chosen subset of 200 of these training structures are used to perform single-point
HF and post-HF calculations.
The HF calculations of the supercell employ a 3$\times$3$\times$3 k-point mesh and
a plane wave basis energy cutoff of 400$\,$eV.
The CCSD and CCSD(T) calculations are performed using the \texttt{Cc4s}
code~\cite{cc4s_website} interfaced to \texttt{VASP}.
In the CC calculations, 10 frozen unoccupied natural orbitals per occupied state
were used.
Furthermore, the post-HF calculations only sample the $\Gamma$-point
of the first Brillouin zone of the supercell.
To correct for the basis-set-incompleteness and finite-size errors
of the correlation energies,
we employ approximate corrections summarized in 
Refs.~\cite{gruber2018applying,irmler2021focal}
also used in recent related work~\cite{herzog2024coupled}.
The WFT training data is generated by selecting $200$ structures randomly from the 
original \texttt{VASP} MD, and computing total energies for each one. The DFT energies 
for said structures are then subtracted, so only the difference of the energies will be 
used in training.

In the case of diamond, ML(DFT$_{\rm{E,F}}$) was trained on the same $200$
structures as $\Delta$ML(WFT) (instead of the full $2000$) as this turned out to be sufficient. 

For the calculations of harmonic phonons and related properties,
we use \texttt{phonopy}~\cite{phonopy-phono3py-JPCM,phonopy-phono3py-JPSJ} with a 
$4\times4\times4$ supercell (of the primitive basis cell, so $128$ atoms);
for the HA-DOS comparison, we use a $24 \times 24 \times 24$ mesh.
To generate primitive basis cells for our structures, we use \texttt{VASPKIT} 
\cite{VASPKIT}.

For the calculation of the VACFs, we used the \texttt{MACE} calculator
for \texttt{ASE}~\cite{ase1,ase2} to create an MD of $30000$ steps at $300\,$K,
based on a $512$ atom ($4\times4\times4$) starting structure; we did this once
using ML(DFT$_{\rm{E,F}}$), and once using $\Delta$ML(CCSD(T)). The resulting MD was
fed into the VACF code taken from Ref.~\cite{vacf}.

\subsubsection{Lithium Hydride}\label{lihcomp}

The procedure for LiH largely followed that for C in Sec.~\ref{ccomp}, with the same 
codes used.
A lattice constant of 4.017$\,$\AA$\,$was used for LiH.
Unlike diamond, interatomic interactions in LiH are dominated by ionic 
bonds, causing long range effects. Thus, a bigger supercell ($2\times2\times2$, $64$ atoms) was necessary in the training data generation, and for ML(DFT$_{\rm{E,F}}$) \emph{all} $2000$ MD steps were used in training. We note that using half the number of training points does not change results, as shown in Sec. 2.1 of the SI. The \texttt{VASP} calculations were done using DFT-LDA. 
As ML(DFT$_{\rm{E}}$) represents our WFT test case, it was trained on $200$ randomly chosen structures.

\section{Data availability}\label{data}
All data generated or analysed during this study are included in this published article and its supplementary information files. The datasets used and/or analysed during the current study are available at~\cite{sidata}.
\section{Code availability}\label{code}
The \texttt{Cc4s} documentation and installation guide can be found at \hyperlink{https://manuals.cc4s.org/user-manual/} {https://manuals.cc4s.org/user-manual/} \\

\noindent The \texttt{MACE} documentation and installation guide can be found at \hyperlink{https://mace-docs.readthedocs.io/en/latest/}{https://mace-docs.readthedocs.io/en/latest/} \\

\noindent The \texttt{QNEP} documentation and installation guide can be found at \hyperlink{https://gpumd.org/nep/index.html}{https://gpumd.org/nep/index.html}

\bibliography{Paper-bibliography}

@misc{si,
title = "Supplementary Information",
author = "Sita Sch{\"o}nbauer and Johanna P. Carbone and Andreas Gr{\"u}neis",
year = "2025"
}

@misc{sidata,
title = "Supporting data for this publication",
author = "Sita Sch{\"o}nbauer and Johanna P. Carbone and Andreas Gr{\"u}neis",
url = "https://github.com/waup-1/MLphononsdata/tree/a2855d3f7111b728ef602a06f2d8ab6ebb236b27",
year = "2025"}

@article{VASPKIT,
title = "VASPKIT: A user-friendly interface facilitating high-throughput computing and analysis using VASP code",
journal = "Computer Physics Communications",
volume = "267",
pages = "108033",
year = "2021",
doi = "https://doi.org/10.1016/j.cpc.2021.108033",
author = "Vei Wang and Nan Xu and Jin-Cheng Liu and Gang Tang and Wen-Tong Geng"
}

@article{vasp1,
  title = "Ab initio molecular dynamics for liquid metals",
  author = "Kresse, G. and Hafner, J.",
  journal = "Phys. Rev. B",
  volume = "47",
  issue = "1",
  pages = "558--561",
  numpages = "0",
  year = "1993",
  month = "Jan",
  publisher = "American Physical Society",
  doi = "10.1103/PhysRevB.47.558",
  url = "https://link.aps.org/doi/10.1103/PhysRevB.47.558"
}

@article{vasp2,
  title = "Efficient iterative schemes for ab initio total-energy calculations using a plane-wave basis set",
  author = "Kresse, G. and Furthm{\"u}ller, J.",
  journal = "Phys. Rev. B",
  volume = "54",
  issue = "16",
  pages = "11169--11186",
  numpages = "0",
  year = "1996",
  month = "Oct",
  publisher = "American Physical Society",
  doi = "10.1103/PhysRevB.54.11169",
  url = "https://link.aps.org/doi/10.1103/PhysRevB.54.11169"
}

@article{vasp3,
title = {Efficiency of ab-initio total energy calculations for metals and semiconductors using a plane-wave basis set},
journal = {Computational Materials Science},
volume = {6},
number = {1},
pages = {15-50},
year = {1996},
issn = {0927-0256},
doi = {https://doi.org/10.1016/0927-0256(96)00008-0},
author = {G. Kresse and J. Furthmüller}
}

@article{phonopy-phono3py-JPCM,
  author  = "Togo, Atsushi and Chaput, Laurent and Tadano, Terumasa and Tanaka, Isao",
  title   = "Implementation strategies in phonopy and phono3py",
  journal = "J. Phys. Condens. Matter",
  volume  = "35",
  number  = "35",
  pages   = "353001",
  year    = "2023",
  doi     = "10.1088/1361-648X/acd831"
}

@article{phonopy-phono3py-JPSJ,
  author  = "Togo, Atsushi",
  title   = "First-principles Phonon Calculations with Phonopy and Phono3py",
  journal = "J. Phys. Soc. Jpn.",
  volume  = "92",
  number  = "1",
  pages   = "012001",
  year    = "2023",
  doi     = "10.7566/JPSJ.92.012001"
}

@article{Batatia2022Design,
  title={The design space of E(3)-equivariant atom-centred interatomic potentials},
  author={Batatia, Ilyes and Batzner, Simon and Kov{\'a}cs, D{\'a}vid P. and Musaelian, Albert and Simm, Gregor N. C. and Drautz, Ralf and Ortner, Christoph and Kozinsky, Boris and Cs{\'a}nyi, G{\'a}bor},
  journal={Nature Machine Intelligence},
  volume={7},
  number={1},
  pages={56--67},
  year={2025},
  publisher={Nature Publishing Group},
  doi={10.1038/s42256-024-00956-x},
  url={https://doi.org/10.1038/s42256-024-00956-x}
}

@inproceedings{Batatia2022mace,
  title={{MACE}: Higher Order Equivariant Message Passing Neural Networks for Fast and Accurate Force Fields},
  author={Ilyes Batatia and David Peter Kovacs and Gregor N. C. Simm and Christoph Ortner and Gabor Csanyi},
  booktitle={Advances in Neural Information Processing Systems},
  editor={Alice H. Oh and Alekh Agarwal and Danielle Belgrave and Kyunghyun Cho},
  year={2022},
  url={https://openreview.net/forum?id=YPpSngE-ZU}
}

@article{lihexp,
title = "The phonon spectra of LiH and LiD from density-functional perturbation theory",
journal = "Solid State Communications",
volume = "98",
number = "3",
pages = "203-207",
year = "1996",
issn = "0038-1098",
doi = "https://doi.org/10.1016/0038-1098(96)00067-1",
url = "https://www.sciencedirect.com/science/article/pii/0038109896000671",
author = "Guido Roma and Carlo M. Bertoni and Stefano Baroni"
}

@article{ase1,
doi = "10.1088/1361-648X/aa680e",
url = "https://dx.doi.org/10.1088/1361-648X/aa680e",
year = "2017",
month = "jun",
publisher = "IOP Publishing",
volume = "29",
number = "27",
pages = "273002",
author = "Ask Hjorth Larsen and Jens Jørgen Mortensen and Jakob Blomqvist and Ivano E Castelli and Rune Christensen and Marcin Dułak and Jesper Friis and Michael N Groves and Bjørk Hammer and Cory Hargus and Eric D Hermes and Paul C Jennings and Peter Bjerre Jensen and James Kermode and John R Kitchin and Esben Leonhard Kolsbjerg and Joseph Kubal and Kristen Kaasbjerg and Steen Lysgaard and Jón Bergmann Maronsson and Tristan Maxson and Thomas Olsen and Lars Pastewka and Andrew Peterson and Carsten Rostgaard and Jakob Schiøtz and Ole Schütt and Mikkel Strange and Kristian S Thygesen and Tejs Vegge and Lasse Vilhelmsen and Michael Walter and Zhenhua Zeng and Karsten W Jacobsen",
title = "The atomic simulation environment—a Python library for working with atoms",
journal = "Journal of Physics: Condensed Matter"
}

@ARTICLE{ase2,

  author="Bahn, S.R. and Jacobsen, K.W.",

  journal="Computing in Science \& Engineering", 

  title="An object-oriented scripting interface to a legacy electronic structure code", 

  year="2002",

  volume="4",

  number="3",

  pages="56-66",

  doi="10.1109/5992.998641"}

@article{cqmcgamma,
  title = "Equation of State and Raman Frequency of Diamond from Quantum Monte Carlo Simulations",
  author = "Maezono, Ryo and Ma, A. and Towler, M. D. and Needs, R. J.",
  journal = "Phys. Rev. Lett.",
  volume = "98",
  issue = "2",
  pages = "025701",
  numpages = "4",
  year = "2007",
  month = "Jan",
  publisher = "American Physical Society",
  doi = "10.1103/PhysRevLett.98.025701",
  url = "https://link.aps.org/doi/10.1103/PhysRevLett.98.025701"
}

@article{cexp,
  title = "Lattice Dynamics of Diamond",
  author = "Warren, J. L. and Yarnell, J. L. and Dolling, G. and Cowley, R. A.",
  journal = "Phys. Rev.",
  volume = "158",
  issue = "3",
  pages = "805--808",
  numpages = "0",
  year = "1967",
  month = "Jun",
  publisher = "American Physical Society",
  doi = "10.1103/PhysRev.158.805",
  url = "https://link.aps.org/doi/10.1103/PhysRev.158.805"
}

@article{cgammaexp,
  title={Properties of diamond under hydrostatic pressures up to 140 GPa},
  author={Occelli, F. and Loubeyre, P. and LeToullec, R.},
  journal={Nature Materials},
  volume={2},
  number={3},
  pages={151--154},
  year={2003},
  publisher={Nature Publishing Group},
  doi={10.1038/nmat831},
  url={https://doi.org/10.1038/nmat831}
}

@unpublished{vacf,
    author = "Huan Wang",
    title = "Velocity-ACF",
    note = "github code",
    url = "https://github.com/LePingKYXK/Velocity-ACF"
}

@article{lihdosexp,
title = {Extraction of the density of phonon states in LiH and NaH},
journal = {Physica B: Condensed Matter},
volume = {350},
number = {1, Supplement },
pages = {E983-E986},
year = {2004},
issn = {0921-4526},
doi = {https://doi.org/10.1016/j.physb.2004.03.271},
url = {https://www.sciencedirect.com/science/article/pii/S0921452604005149},
author = {D Colognesi and A.J Ramirez-Cuesta and M Zoppi and R Senesi and T Abdul-Redah}
}

@article{lihanh,
  title = {Anharmonicity in Thermal Insulators: An Analysis from First Principles},
  author = {Knoop, Florian and Purcell, Thomas A. R. and Scheffler, Matthias and Carbogno, Christian},
  journal = {Phys. Rev. Lett.},
  volume = {130},
  issue = {23},
  pages = {236301},
  numpages = {6},
  year = {2023},
  month = {Jun},
  publisher = {American Physical Society},
  doi = {10.1103/PhysRevLett.130.236301},
  url = {https://link.aps.org/doi/10.1103/PhysRevLett.130.236301}
}

@article{Raghavachari1989,
title = {A fifth-order perturbation comparison of electron correlation theories},
journal = {Chemical Physics Letters},
volume = {157},
number = {6},
pages = {479-483},
year = {1989},
issn = {0009-2614},
doi = {https://doi.org/10.1016/S0009-2614(89)87395-6},
url = {https://www.sciencedirect.com/science/article/pii/S0009261489873956},
author = {Krishnan Raghavachari and Gary W. Trucks and John A. Pople and Martin Head-Gordon}
}

@article{deltacc1,
author={Bogojeski, Mihail and Vogt-Maranto, Leslie and Tuckerman, Mark E. and M{\"u}ller, Klaus-Robert and Burke, Kieron},
title={Quantum chemical accuracy from density functional approximations via machine learning},
journal={Nature Communications},
year={2020},
month={Oct},
day={16},
volume={11},
number={1},
pages={5223},
issn={2041-1723},
doi={10.1038/s41467-020-19093-1},
url={https://doi.org/10.1038/s41467-020-19093-1}
}

@article{deltacc2,
author ="Qu, Chen and Yu, Qi and Conte, Riccardo and Houston, Paul L. and Nandi, Apurba and Bomwan, Joel M.",
title  ="A $\Delta$-machine learning approach for force fields{,} illustrated by a CCSD(T) 4-body correction to the MB-pol water potential",
journal  ="Digital Discovery",
year  ="2022",
volume  ="1",
issue  ="5",
pages  ="658-664",
publisher  ="RSC",
doi  ="10.1039/D2DD00057A",
url  ="http://dx.doi.org/10.1039/D2DD00057A"}

@Article{mlmd1,
author ="Gastegger, Michael and Behler, Jörg and Marquetand, Philipp",
title  ="Machine learning molecular dynamics for the simulation of infrared spectra",
journal  ="Chem. Sci.",
year  ="2017",
volume  ="8",
issue  ="10",
pages  ="6924-6935",
publisher  ="The Royal Society of Chemistry",
doi  ="10.1039/C7SC02267K",
url  ="http://dx.doi.org/10.1039/C7SC02267K"}

@article{mlmd2gpr,
author = {John, S. T. and Csányi, Gábor},
title = {Many-Body Coarse-Grained Interactions Using Gaussian Approximation Potentials},
journal = {The Journal of Physical Chemistry B},
volume = {121},
number = {48},
pages = {10934-10949},
year = {2017},
doi = {10.1021/acs.jpcb.7b09636},
URL = { 
        https://doi.org/10.1021/acs.jpcb.7b09636
}
}

@article{mlmd3,
author = {Stefan Chmiela  and Alexandre Tkatchenko  and Huziel E. Sauceda  and Igor Poltavsky  and Kristof T. Schütt  and Klaus-Robert Müller },
title = {Machine learning of accurate energy-conserving molecular force fields},
journal = {Science Advances},
volume = {3},
number = {5},
pages = {e1603015},
year = {2017},
doi = {10.1126/sciadv.1603015},
URL = {https://www.science.org/doi/abs/10.1126/sciadv.1603015}}

@article{mlmd4code,
title = {sGDML: Constructing accurate and data efficient molecular force fields using machine learning},
journal = {Computer Physics Communications},
volume = {240},
pages = {38-45},
year = {2019},
issn = {0010-4655},
doi = {https://doi.org/10.1016/j.cpc.2019.02.007},
url = {https://www.sciencedirect.com/science/article/pii/S0010465519300591},
author = {Stefan Chmiela and Huziel E. Sauceda and Igor Poltavsky and Klaus-Robert Müller and Alexandre Tkatchenko}
}

@article{mlmd5symm,
  title = {Deep Potential Molecular Dynamics: A Scalable Model with the Accuracy of Quantum Mechanics},
  author = {Zhang, Linfeng and Han, Jiequn and Wang, Han and Car, Roberto and E, Weinan},
  journal = {Phys. Rev. Lett.},
  volume = {120},
  issue = {14},
  pages = {143001},
  numpages = {6},
  year = {2018},
  month = {Apr},
  publisher = {American Physical Society},
  doi = {10.1103/PhysRevLett.120.143001},
  url = {https://link.aps.org/doi/10.1103/PhysRevLett.120.143001}
}

@Article{transfer,
author={Smith, Justin S.
and Nebgen, Benjamin T.
and Zubatyuk, Roman
and Lubbers, Nicholas
and Devereux, Christian
and Barros, Kipton
and Tretiak, Sergei
and Isayev, Olexandr
and Roitberg, Adrian E.},
title={Approaching coupled cluster accuracy with a general-purpose neural network potential through transfer learning},
journal={Nature Communications},
year={2019},
month={Jul},
day={01},
volume={10},
number={1},
pages={2903},
issn={2041-1723},
doi={10.1038/s41467-019-10827-4},
url={https://doi.org/10.1038/s41467-019-10827-4}
}

@Article{paraelectricity,
author={Kim, Kyoung-Min
and Chung, Suk Bum},
title={Phonon-mediated spin transport in quantum paraelectric metals},
journal={npj Quantum Materials},
year={2024},
month={Jul},
day={04},
volume={9},
number={1},
pages={51},
issn={2397-4648},
doi={10.1038/s41535-024-00662-2},
url={https://doi.org/10.1038/s41535-024-00662-2}
}

@article{superconductivity,
author = {John Bardeen },
title = {Electron-Phonon Interactions and Superconductivity},
journal = {Science},
volume = {181},
number = {4106},
pages = {1209-1214},
year = {1973},
doi = {10.1126/science.181.4106.1209},
URL = {https://www.science.org/doi/abs/10.1126/science.181.4106.1209}}

@article{Bartlett2007,
  title = {Coupled-cluster theory in quantum chemistry},
  author = {Bartlett, Rodney J. and Musia\l{}, Monika},
  journal = {Rev. Mod. Phys.},
  volume = {79},
  issue = {1},
  pages = {291--352},
  numpages = {0},
  year = {2007},
  month = {Feb},
  publisher = {American Physical Society},
  doi = {10.1103/RevModPhys.79.291},
  url = {https://link.aps.org/doi/10.1103/RevModPhys.79.291}
}

@article{Daru2022,
  title = {Coupled Cluster Molecular Dynamics of Condensed Phase Systems Enabled by Machine Learning Potentials: Liquid Water Benchmark},
  author = {Daru, J\'anos and Forbert, Harald and Behler, J\"org and Marx, Dominik},
  journal = {Phys. Rev. Lett.},
  volume = {129},
  issue = {22},
  pages = {226001},
  numpages = {6},
  year = {2022},
  month = {Nov},
  publisher = {American Physical Society},
  doi = {10.1103/PhysRevLett.129.226001},
  url = {https://link.aps.org/doi/10.1103/PhysRevLett.129.226001}
}

@article{chen2023data,
author = {Chen, Michael S. and Lee, Joonho and Ye, Hong-Zhou and Berkelbach, Timothy C. and Reichman, David R. and Markland, Thomas E.},
title = {Data-Efficient Machine Learning Potentials from Transfer Learning of Periodic Correlated Electronic Structure Methods: Liquid Water at AFQMC, CCSD, and CCSD(T) Accuracy},
journal = {Journal of Chemical Theory and Computation},
volume = {19},
number = {14},
pages = {4510-4519},
year = {2023},
doi = {10.1021/acs.jctc.2c01203},
URL = {https://doi.org/10.1021/acs.jctc.2c01203}}

@article{herzog2024coupled,
  title={Coupled cluster finite temperature simulations of periodic materials via machine learning},
  author={Herzog, Basile and Gallo, Alejandro and Hummel, Felix and Badawi, Michael and Bu{\v{c}}ko, Tom{\'a}{\v{s}} and Leb{\`e}gue, S{\'e}bastien and Gr{\"u}neis, Andreas and Rocca, Dario},
  journal={npj Computational Materials},
  volume={10},
  number={1},
  pages={68},
  year={2024},
  publisher={Nature Publishing Group UK London},
  doi = {https://doi.org/10.1038/s41524-024-01249-y}
}

@article{Unke2021,
author = {Unke, Oliver T. and Chmiela, Stefan and Sauceda, Huziel E. and Gastegger, Michael and Poltavsky, Igor and Sch{\"u}tt, Kristof T. and Tkatchenko, Alexandre and M{\"u}ller, Klaus-Robert},
title = {Machine Learning Force Fields},
journal = {Chemical Reviews},
volume = {121},
number = {16},
pages = {10142-10186},
year = {2021},
doi = {10.1021/acs.chemrev.0c01111},
URL = {
        https://doi.org/10.1021/acs.chemrev.0c01111
}
}

@article{Ruth2023,
author = {Ruth, Marcel and Gerbig, Dennis and Schreiner, Peter R.},
title = {Machine Learning for Bridging the Gap between Density Functional Theory and Coupled Cluster Energies},
journal = {Journal of Chemical Theory and Computation},
volume = {19},
number = {15},
pages = {4912-4920},
year = {2023},
doi = {10.1021/acs.jctc.3c00274},
URL = {
        https://doi.org/10.1021/acs.jctc.3c00274
}
}

@article{Behler2007,
  title = {Generalized Neural-Network Representation of High-Dimensional Potential-Energy Surfaces},
  author = {Behler, J\"org and Parrinello, Michele},
  journal = {Phys. Rev. Lett.},
  volume = {98},
  issue = {14},
  pages = {146401},
  numpages = {4},
  year = {2007},
  month = {Apr},
  publisher = {American Physical Society},
  doi = {10.1103/PhysRevLett.98.146401},
  url = {https://link.aps.org/doi/10.1103/PhysRevLett.98.146401}
}

@article{Bartok2010,
  title = {Gaussian Approximation Potentials: The Accuracy of Quantum Mechanics, without the Electrons},
  author = {Bart\'ok, Albert P. and Payne, Mike C. and Kondor, Risi and Cs\'anyi, G\'abor},
  journal = {Phys. Rev. Lett.},
  volume = {104},
  issue = {13},
  pages = {136403},
  numpages = {4},
  year = {2010},
  month = {Apr},
  publisher = {American Physical Society},
  doi = {10.1103/PhysRevLett.104.136403},
  url = {https://link.aps.org/doi/10.1103/PhysRevLett.104.136403}
}

@misc{cc4s_website,
  title     = {{Cc4s}},
  year      = {2025},
  url       = {https://manuals.cc4s.org/user-manual}}

@article{gruber2018applying,
   title = {Applying the Coupled-Cluster Ansatz to Solids and Surfaces in the Thermodynamic Limit},
  author = {Gruber, Thomas and Liao, Ke and Tsatsoulis, Theodoros and Hummel, Felix and Gr\"uneis, Andreas},
  journal = {Phys. Rev. X},
  volume = {8},
  issue = {2},
  pages = {021043},
  numpages = {8},
  year = {2018},
  month = {May},
  publisher = {American Physical Society},
  doi = {10.1103/PhysRevX.8.021043},
  url = {https://link.aps.org/doi/10.1103/PhysRevX.8.021043}
}

@article{irmler2021focal,
    author = {Irmler, Andreas and Gallo, Alejandro and Grüneis, Andreas},
    title = {Focal-point approach with pair-specific cusp correction for coupled-cluster theory},
    journal = {The Journal of Chemical Physics},
    volume = {154},
    number = {23},
    pages = {234103},
    year = {2021},
    month = {06},
    issn = {0021-9606},
    doi = {10.1063/5.0050054},
    url = {https://doi.org/10.1063/5.0050054},
}

@article{ranalli2023temperature,
author = {Ranalli, Luigi and Verdi, Carla and Monacelli, Lorenzo and Kresse, Georg and Calandra, Matteo and Franchini, Cesare},
title = {Temperature-Dependent Anharmonic Phonons in Quantum Paraelectric KTaO3 by First Principles and Machine-Learned Force Fields},
journal = {Advanced Quantum Technologies},
volume = {6},
number = {4},
pages = {2200131},
doi = {https://doi.org/10.1002/qute.202200131},
url = {https://advanced.onlinelibrary.wiley.com/doi/abs/10.1002/qute.202200131},
year = {2023}
}

@article{kulic2000interplay,
  title={Interplay of electron--phonon interaction and strong correlations: the possible way to high-temperature superconductivity},
  author={Kuli{\'c}, Miodrag L},
  journal={Physics Reports},
  volume={338},
  number={1-2},
  pages={1--264},
  year={2000},
  publisher={Elsevier},
  doi={10.1016/S0370-1573(00)00008-9}
}

@article{BARRERA2005119,
title = {LDA or GGA? A combined experimental inelastic neutron scattering and ab initio lattice dynamics study of alkali metal hydrides},
journal = {Chemical Physics},
volume = {317},
number = {2},
pages = {119-129},
year = {2005},
issn = {0301-0104},
doi = {https://doi.org/10.1016/j.chemphys.2005.04.027},
url = {https://www.sciencedirect.com/science/article/pii/S0301010405001746},
author = {G.D. Barrera and D. Colognesi and P.C.H. Mitchell and A.J. Ramirez-Cuesta}
}

@article{Gonze1997,
  title = {Dynamical matrices, Born effective charges, dielectric permittivity tensors, and interatomic force constants from density-functional perturbation theory},
  author = {Gonze, Xavier and Lee, Changyol},
  journal = {Phys. Rev. B},
  volume = {55},
  issue = {16},
  pages = {10355--10368},
  numpages = {0},
  year = {1997},
  month = {Apr},
  publisher = {American Physical Society},
  doi = {10.1103/PhysRevB.55.10355},
  url = {https://link.aps.org/doi/10.1103/PhysRevB.55.10355}
}

@article{transfer2,
author = {K{\"a}ser, Silvan and Richardson, Jeremy O. and Meuwly, Markus},
title = {Transfer Learning for Predictive Molecular Simulations: Data-Efficient Potential Energy Surfaces at CCSD(T) Accuracy},
journal = {Journal of Chemical Theory and Computation},
volume = {0},
number = {0},
pages = {null},
year = {0},
doi = {10.1021/acs.jctc.5c00523},
URL = {         https://doi.org/10.1021/acs.jctc.5c00523} }

@misc{transfermariia,
      title={Fine-tuning foundation models of materials interatomic potentials with frozen transfer learning}, 
      author={Mariia Radova and Wojciech G. Stark and Connor S. Allen and Reinhard J. Maurer and Albert P. Bartók},
      year={2025},
      eprint={2502.15582},
      archivePrefix={arXiv},
      primaryClass={cond-mat.mtrl-sci},
      url={https://arxiv.org/abs/2502.15582}, 
}

@article{behler,
author = {Behler, J{\"o}rg},
title = {Four Generations of High-Dimensional Neural Network Potentials},
journal = {Chemical Reviews},
volume = {121},
number = {16},
pages = {10037-10072},
year = {2021},
doi = {10.1021/acs.chemrev.0c00868}
}

@article{qnep1,
author = {Xu, Ke and Bu, Hekai and Pan, Shuning and Lindgren, Eric and Wu, Yongchao and Wang, Yong and Liu, Jiahui and Song, Keke and Xu, Bin and Li, Yifan and Hainer, Tobias and Svensson, Lucas and Wiktor, Julia and Zhao, Rui and Huang, Hongfu and Qian, Cheng and Zhang, Shuo and Zeng, Zezhu and Zhang, Bohan and Tang, Benrui and Xiao, Yang and Yan, Zihan and Shi, Jiuyang and Liang, Zhixin and Wang, Junjie and Liang, Ting and Cao, Shuo and Wang, Yanzhou and Ying, Penghua and Xu, Nan and Chen, Chengbing and Zhang, Yuwen and Chen, Zherui and Wu, Xin and Jiang, Wenwu and Berger, Esme and Li, Yanlong and Chen, Shunda and Gabourie, Alexander J. and Dong, Haikuan and Xiong, Shiyun and Wei, Ning and Chen, Yue and Xu, Jianbin and Ding, Feng and Sun, Zhimei and Ala-Nissila, Tapio and Harju, Ari and Zheng, Jincheng and Guan, Pengfei and Erhart, Paul and Sun, Jian and Ouyang, Wengen and Su, Yanjing and Fan, Zheyong},
title = {GPUMD 4.0: A high-performance molecular dynamics package for versatile materials simulations with machine-learned potentials},
journal = {Materials Genome Engineering Advances},
volume = {3},
number = {3},
pages = {e70028},
doi = {https://doi.org/10.1002/mgea.70028},
year = {2025}
}

@article{qnep2,
  title = {Neuroevolution machine learning potentials: Combining high accuracy and low cost in atomistic simulations and application to heat transport},
  author = {Fan, Zheyong and Zeng, Zezhu and Zhang, Cunzhi and Wang, Yanzhou and Song, Keke and Dong, Haikuan and Chen, Yue and Ala-Nissila, Tapio},
  journal = {Phys. Rev. B},
  volume = {104},
  issue = {10},
  pages = {104309},
  numpages = {15},
  year = {2021},
  month = {Sep},
  publisher = {American Physical Society},
  doi = {10.1103/PhysRevB.104.104309},
  url = {https://link.aps.org/doi/10.1103/PhysRevB.104.104309}
}

@misc{qnep3,
      title={qNEP: A highly efficient neuroevolution potential with dynamic charges for large-scale atomistic simulations}, 
      author={Zheyong Fan and Benrui Tang and Esmée Berger and Ethan Berger and Erik Fransson and Ke Xu and Zihan Yan and Zhoulin Liu and Zichen Song and Haikuan Dong and Shunda Chen and Lei Li and Ziliang Wang and Yizhou Zhu and Julia Wiktor and Paul Erhart},
      year={2026},
      eprint={2601.19034},
      archivePrefix={arXiv},
      primaryClass={physics.comp-ph},
      url={https://arxiv.org/abs/2601.19034}, 
}

\section{Acknowledgements}
Support and funding from the European Research Council (ERC) (Grant Agreement No 101087184) is gratefully acknowledged.
This research was funded in part by the Austrian Science Fund (FWF) 10.55776/COE5 MECS. For Open Access purposes, the author has applied a CC BY public copyright license to any author accepted manuscript version arising from this submission.
The computational results presented have been achieved using the Vienna Scientific Cluster (VSC).
\section{Author contributions}
S.S. performed all calculations, implemented the workflows used for the present work and analyzed the data. J.P.C. and S.S. produced the initial draft of the manuscript. F.V.E. and F.L. consulted on the machine learning methods. J.P.C. and A.G. supervised the work and
contributed to writing the final manuscript.

\section{Competing interests}
The authors declare no competing interests.

\end{multicols}

\end{document}